\documentclass{article}

\usepackage{dcolumn}
\usepackage{amsmath}
\usepackage{amssymb}
\usepackage{bm}

\usepackage{graphicx}
\usepackage{subfig}

\usepackage{hyperref}

\usepackage{color}
\usepackage{ulem}

\usepackage[flushleft]{threeparttable}

\usepackage{authblk}

\title{Development of ion-beam sputtered silicon nitride thin films for low-noise mirror coatings of gravitational-wave detectors}

\author[1,2]{A. Amato}
\affil[1]{Maastricht University, Minderbroedersberg 4-6, 6211 LK Maastricht, The Netherlands}
\affil[2]{Nikhef, Science Park 105, 1098 XG Amsterdam, The Netherlands}
\author[3]{M. Bazzan}
\affil[3]{Universit\`a di Padova, Dipartimento di Fisica e Astronomia, Via Marzolo 8, 35131 Padova, Italy}
\author[4]{G. Cagnoli}
\affil[4]{Universit\'e de Lyon, Universit\'e Claude Bernard Lyon 1, CNRS, Institut Lumi\`ere Mati\`ere, 69100 Villeurbanne, France}
\author[5]{M. Canepa}
\affil[5]{OptMatLab, Dipartimento di Fisica, Università di Genova, Via Dodecaneso 33, 16146
Genova, Italy}
\author[6]{M. Coulon}
\affil[6]{Laboratoire des Mat\'eriaux Avanc\'es - IP2I, CNRS, Universit\'{e} de Lyon, Universit\'{e} Claude Bernard Lyon 1, 69100 Villeurbanne, France}
\author[6]{J. Degallaix}
\author[7]{N. Demos}
\affil[7]{Massachusetts Institute of Technology, 185 Albany Street NW22-295, Cambridge,
Massachusetts 02139, USA}
\author[8,9] {A. Di Michele}
\affil[8]{Dipartimento di Fisica e Geologia, Universit\`a di Perugia, 06100 Perugia, Italy}
\affil[9]{INFN Sezione di Perugia, 06100 Perugia, Italy}
\author[7]{M. Evans}
\author[10,11]{F. Fabrizi}
\affil[10]{Universit\`a degli Studi di Urbino “Carlo Bo”, 61029 Urbino, Italy}
\affil[11]{INFN Sezione di Firenze, 50019 Sesto Fiorentino, Firenze, Italy}
\author[3]{G. Favaro}
\author[6]{D. Forest}
\author[7]{S. Gras}
\author[6]{D. Hofman}
\author[12]{A. Lema\^itre}
\affil[12]{Laboratoire Navier, \'Ecole des Ponts, Universit\'e Gustave Eiffel, CNRS, 77420 Marne-la-Vall\'ee, France}
\author[3,13]{G. Maggioni}
\affil[13]{INFN Laboratori Nazionali di Legnaro, Viale dell'Universit\`a 2, 35020 Legnaro, Padova, Italy }
\author[5,14]{M. Magnozzi}
\affil[14]{INFN Sezione di Genova, Via Dodecaneso 33, 16146 Genova, Italy}
\author[4]{V. Martinez}
\author[6]{L. Mereni}
\author[6]{C. Michel}
\author[3]{V. Milotti}
\author[10,11]{M. Montani}
\author[15,16]{A. Paolone}
\affil[15]{Istituto dei Sistemi Complessi, CNR, Piazzale A. Moro 5, 00185 Roma, Italy}
\affil[16]{INFN Sezione di Roma, Piazzale A. Moro 5, 00185 Roma, Italy}
\author[4]{A. Pereira}
\author[10,11]{F. Piergiovanni}
\author[17,18]{V. Pierro}
\affil[17]{Dipartimento di Ingegneria (DING), Universit\`a del Sannio, 82100 Benevento, Italy}
\affil[18]{INFN Sezione di Napoli Gruppo Collegato di Salerno, Complesso Universitario di Monte S. Angelo, 80126 Napoli, Italy}
\author[6]{L. Pinard}
\author[18]{I. M. Pinto}
\author[16,19]{E. Placidi}
\affil[19]{Sapienza Universit\`a di Roma, Dipartimento di Fisica, Piazzale A. Moro 5, 00185 Roma, Italy}
\author[5]{S. Samandari}
\author[6]{B. Sassolas}
\author[12,20]{N. Shcheblanov}
\affil[20]{Laboratoire Mod\'elisation et Simulation Multi-\'Echelle, Universit\'e Gustave Eiffel, CNRS, Universit\'e Paris Est Cr\'eteil, 77454 Marne-la-Vall\'ee, France}
\author[6]{J. Teillon}
\author[21]{I. Vickridge}
\affil[21]{Sorbonne Universit\'e, CNRS, Institut des NanoSciences de Paris (INSP), SAFIR, 75005 Paris, France}
\author[6]{M. Granata\thanks{m.granata@lma.in2p3.fr}}

\date{\today}

\begin{document}

\maketitle

\begin{abstract}
Brownian thermal noise of thin-film coatings is a fundamental limit for high-precision experiments based on optical resonators such as gravitational-wave interferometers. Here we present the results of a research activity aiming to develop lower-noise ion-beam sputtered silicon nitride thin films compliant with the very stringent requirements on optical loss of gravitational-wave interferometers.

In order to test the hypothesis of a correlation between the synthesis conditions of the films and their elemental composition and optical and mechanical properties, we varied the voltage, current intensity and composition of the sputtering ion beam, and we performed a broad campaign of characterizations. While the refractive index was found to monotonically depend on the beam voltage and linearly vary with the N/Si ratio, the optical absorption appeared to be strongly sensitive to other factors, as yet unidentified. However, by systematically varying the deposition parameters, an optimal working point was found. Thus we show that the loss angle and extinction coefficient of our thin films can be as low as $(1.0 \pm 0.1) \times 10^{-4}$ rad at $\sim$2.8 kHz and $(6.4 \pm 0.2) \times 10^{-6}$ at 1064 nm, respectively, after thermal treatment at 900 $^{\circ}$C. To the best of our knowledge, such loss angle value is the lowest ever measured on this class of thin films.

We then used our silicon nitride thin films to design and produce a multi-material mirror coating showing a thermal noise amplitude of $(10.3 \pm 0.2) \times 10^{-18}$ m Hz$^{-1/2}$ at 100 Hz, which is 25\% lower than in current mirror coatings of the Advanced LIGO and Advanced Virgo interferometers, and an optical absorption as low as $(1.6 \pm 0.5)$ parts per million at 1064 nm.
\end{abstract}

\section{Introduction}
The ground-based gravitational-wave detectors LIGO \cite{aLIGO}, Virgo \cite{AdV} and KAGRA \cite{KAGRA} are km-scale Michelson interferometers with a Fabry-Perot cavity in each arm \cite{saulson17}. Those cavities consist of two opposing large and massive suspended mirrors, which play the role of field probes for gravitational waves of astrophysical origin. Their high-reflection mirror coatings must satisfy very stringent specifications imposed by extreme sensitivity requirements, that is, very low optical losses \cite{degallaix19,pinard17} and low Brownian thermal noise \cite{levin98,saulson90} at the same time.

The current mirror coatings of LIGO, Virgo and KAGRA are thickness-optimized Bragg reflectors \cite{villar10} deposited by ion-beam sputtering (IBS) at the Laboratoire des Mat\'{e}riaux Avanc\'{e}s (LMA). They are made of stacked alternate thin layers of tantalum pentoxide (Ta$_2$O$_5$, also known as {\it tantala}, high refractive index), either plain or mixed with titanium dioxide (TiO$_2$ or {\it titania}, high refractive index), and silicon dioxide (SiO$_2$ or {\it silica}, low refractive index) \cite{granata20b,mori23}. These coatings are already fully compliant with the optical loss requirements of present gravitational-wave interferometers, but still show a thermal noise amplitude which is expected to be a severe sensitivity limitation \cite{aLIGO,AdV,ET}. For reference, the current Bragg reflectors of LIGO and Virgo feature an optical loss due to absorption and scattering of 1 and 5 parts per million (ppm) \cite{degallaix19,pinard17}, respectively, and a thermal noise amplitude of $(13.7 \pm 0.3) \times 10^{-18}$ m Hz$^{-1/2}$ (see Section \ref{SEC_HR}).

Indeed, according to the fluctuation-dissipation theorem \cite{callen52}, the power spectral density of Brownian noise fluctuations at the coatings' surface is proportional to the rate of elastic energy dissipation within the thin coatings' layers. This rate, quantified by the loss angle $\varphi$, must hence be reduced as much as possible to decrease coating thermal noise. At room temperature, the loss angle of tantala and tantala-titania layers is about an order of magnitude higher than that of silica layers \cite{granata20b,penn03}, making it the dominant source of dissipation. So, since the early 2000s, a considerable research effort has been committed to developing alternative high-index coating materials of lower loss angle \cite{granata20a,mcghee23,vajente21}.

Because of their high refractive index \cite{lee10} and lower loss angle \cite{liu07}, silicon nitride (SiN$_x$) thin films are a promising candidate for future mirror coatings of gravitational-wave interferometers, whether at ambient or cryogenic temperature \cite{granata20a,pan18,steinlechner17}. However, despite recent progress, their optical absorption still remains too high \cite{granata20a,wallace24,tsai22}.

In this work we present the results of our research and development activity on IBS silicon nitride thin films, carried out over the past several years to simultaneously minimize their loss angle and optical absorption and thus enable their use in gravitational-wave interferometers.

\section{Methods}
This section describes the synthesis process of our thin-film samples as well as the techniques used to determine their properties.

\subsection{Samples}
In IBS, a beam of accelerated ions is used to sputter a solid target material into a vapor phase which then condenses into a thin film onto a substrate. Indeed, the deposition of stoichiometric IBS  SiN$_x$ thin films (where ideally $x$ = N/Si = 1.33) depends on multiple  parameters, the most critical ones being the nature, energy and current of the sputtering particles \cite{lee10,lee08,huang97,lambrinos96,ray94,bouchier86,bosseboeuf85,grill83,bouchier83,erler80,weissmantel76}, the deposition rate \cite{bouchier83,weissmantel76}, and the angles between the ion beam and the target and between the target and the substrate holder \cite{bouchier91,burdovitsin83}.

The thin films used in this work were deposited in a custom-developed facility at the LMA. We used several sets of deposition parameters, in order to test the hypothesis of a correlation between the synthesis conditions of the thin-film samples and their composition and optical and mechanical properties. Hence we varied the kinetic energy and current intensity of the sputtering particles, as well as the elemental composition of the ion beam impinging on the target. At the same time, for every deposition run, we kept the beam/target and target/substrate holder angles constant and equal to 45 deg \cite{bouchier91,burdovitsin83}, the working pressure during deposition was of the order of $10^{-4}$ mbar, the substrates were heated up to a temperature between 100 and 150 $^{\circ}$C while rotating at a speed of 3 rpm. Prior to deposition, the base pressure inside the chamber was lower than $2 \times 10^{-7}$ mbar.

\begin{figure}
\includegraphics[width=0.9\textwidth]{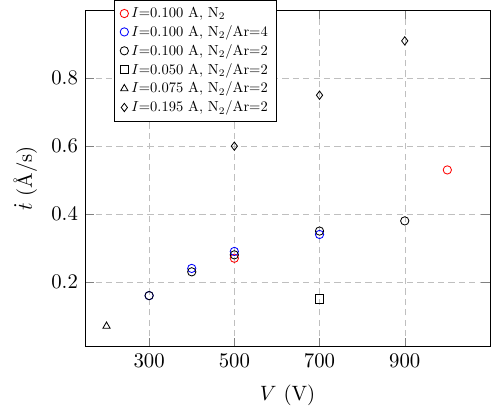}
\caption{\label{FIG_v}Average deposition rate $\dot{t}$ of IBS SiN$_x$ thin-film samples, as a function of ion beam settings: voltage $V$ and current $I$, and ratio N$_2$/Ar of gas mass flows to the source. Uncertainties on $\dot{t}$ are $\leq 1$\% (not shown).}
\centering
\end{figure}

Coating samples were 50 to 500 nm thick and deposited on different substrates, according to the specific conditions required by each planned characterization measurement to yield optimal results. Table \ref{TAB_samples} and Fig.\ref{FIG_v} present the deposition parameters used to deposit the thin-film samples studied in this work (for conciseness, only a representative subset of samples is reported). Those parameters are the beam voltage $V$ and current $I$, which determine the kinetic energy and flow rate of sputtering ions, respectively, the mass flows of argon and nitrogen gases fed to the ion beam source, and the average rate $\dot{t}$ of film thickness growth (also called average deposition rate).
\begin{table}
\centering
\caption{Relevant deposition parameters of IBS SiN$_x$ thin-film samples: ion beam voltage $V$ and current $I$, Ar and N$_2$ gas mass flows fed to the ion source, and average deposition rate $\dot{t}$. Uncertainties on $\dot{t}$ are $\leq 1$\% (not shown).}
\label{TAB_samples}
\begin{tabular}{c|cccccc}
\hline \hline
Sample & $V$ (V) & $I$ (mA) & Ar (sccm) & N$_2$ (sccm) & $\dot{t}$ (\r{A}/s) & Substrates\\ \hline
20043 & 1000 & 100 & 0 & 10 & 0.53 & SiO$_2$\\
21012 & 400 & 100 & 5 & 10 & 0.23 & SiO$_2$\\
21013 & 500 & 100 & 5 & 10 & 0.28 & SiO$_2$\\
21014 & 300 & 100 & 5 & 10 & 0.16 & SiO$_2$\\
21016 & 200 & 75 & 5 & 10 & 0.07 & SiO$_2$\\
21017 & 400 & 100 & 3 & 12 & 0.24 & SiO$_2$\\
21018 & 500 & 100 & 3 & 12 & 0.29 & SiO$_2$\\
21019 & 300 & 100 & 3 & 12 & 0.16 & SiO$_2$\\
21020 & 700 & 100 & 3 & 12 & 0.34 & SiO$_2$\\
21021 & 700 & 100 & 5 & 10 & 0.35 & SiO$_2$\\
21022 & 900 & 100 & 5 & 10 & 0.38 & SiO$_2$\\
21023 & 700 & 50 & 5 & 10 & 0.15 & SiO$_2$\\
21025 & 700 & 195 & 5 & 10 & 0.75 & SiO$_2$\\
21045 & 700 & 195 & 5 & 10 & 0.56 & Si\\
21048 & 700 & 195 & 5 & 10 & 0.56 & Si\\
22013 & 900 & 195 & 5 & 10 & 0.91 & SiO$_2$\\
22014 & 500 & 195 & 5 & 10 & 0.60 & SiO$_2$\\
22015 & 500 & 100 & 0 & 10 & 0.27 & SiO$_2$\\
22020 & 700 & 195 & 5 & 10 & 0.83 & SiO$_2$, C\\
22022 & 500 & 195 & 5 & 10 & 0.60 & SiO$_2$, C\\
22023 & 900 & 195 & 5 & 10 & 0.91 & SiO$_2$, C\\
22027 & 700 & 195 & 5 & 10 & 0.83 & Si\\
22028 & 700 & 195 & 5 & 10 & 0.83 & Si\\
23017 & 700 & 195 & 5 & 10 & 0.83 & SiO$_2$, C\\
23018 & 700 & 195 & 5 & 10 & 0.53 & SiO$_2$, C\\
23028 & 700 & 195 & 5 & 10 & 0.83 & SiO$_2$, Si\\
c21013 & 700 & 100 & 5 & 10 & 0.25 & SiO$_2$\\
c21015 & 1000 & 75 & 0 & 10 & 0.14 & SiO$_2$\\
d2008 & 700 & 195 & 5 & 10 & 0.80 & SiO$_2$\\
\hline \hline
\end{tabular}
\end{table}

Figure \ref{FIG_v} shows that, as previously observed \cite{huang97,ray94}, the deposition rate increases steadily with the ion beam voltage $V$ and current $I$. We obtained average deposition rates ranging from $\sim0.1$ \r{A}/s to $\sim0.9$ \r{A}/s, which compares to values found in the literature for stoichiometric Si$_3$N$_4$ thin films (0.17-0.5 \r{A}/s) \cite{huang97,ray94,erler80,weissmantel76}.

\subsection{Annealing}
High-temperature thermal treatment, or {\it annealing}, is known to generally minimize the optical absorption and loss angle of thin-film coatings \cite{granata20b,fazio20}, hence we studied the effect of annealing on the properties of our specific samples.

Selected samples were thermally treated at high temperature ($T_a \geq 850$ $^\circ$C), in air or in vacuum ($p_a < 10^{-6}$ mbar), and characterized before and after treatment. Unless specified otherwise, annealing time of 10 hours and controlled heating and cooling rates of 100 $^\circ$C/hour were applied.

\subsection{Optical properties}
The refractive index $n$ and thickness $t$ of the samples were evaluated by fitting experimental photometric spectra. For selected samples, independent ellipsometric measurements were also performed over a broader spectral range. The analysis of ellipsometric spectra also allowed us to estimate the energy gap of the samples.

\subsubsection{Spectroscopic photometry}
Samples were deposited on NEGS1 fused-silica substrates (25.4 mm diameter, 6 mm thickness), then measured for transmittance in the 400–1400 nm wavelength range, at normal incidence, using a Cary 7000 spectroscopic photometer by Agilent.

The refractive index $n$ and thickness $t$ of the samples were first evaluated using an envelope method \cite{cisneros98}, then those results were used as initial values in a numerical least-square regression analysis where the adjustable parameters were $t$ and the $B_i$, $C_i$ coefficients of the Sellmeier dispersion equation,
\begin{equation}
n^2 = 1 + \sum_{i=1}^{3} \frac{B_i \lambda^2}{\lambda^2 - C_i}\ ,
\end{equation}
with $\lambda$ being the wavelength.

\subsubsection{Spectroscopic ellipsometry}
We used a J. A. Woollam, Inc. variable-angle spectroscopic ellipsometer (VASE) to characterize a set of selected samples in the 190-2500 nm wavelength range.

Samples were characterized by measuring the amplitude ratio $\Psi$ and phase difference $\Delta$ of the p- and s-polarized light reflected at their surface \cite{fujiwara07}, hence deposited on single-side polished (100) silicon wafers (50.8 or 76.2 mm diameter, 0.5 mm thickness) to suppress spurious reflections from their rear surface. To maximize the response of the instruments, $\Psi$ and $\Delta$ spectra were acquired at incidence angles of 60, 65 and 70 deg, close to the samples' Brewster angle.

The refractive index $n$, extinction coefficient $k$, thickness $t$ and energy gap $E_g$ of the samples were derived by fitting the acquired spectra with a Cody-Lorentz optical model \cite{ferlauto02}, using the data analysis software WVASE. The optical response of the bare wafers had been characterized with prior specific measurements and analyses. Further details about our analysis are available elsewhere \cite{amato19, amato20}.

\subsubsection{Photo-thermal deflection (PTD)}
Optical absorption at 1064 nm was measured using a custom-developed experimental setup based on PTD \cite{boccara80}, with a sensitivity of less than 0.5 parts per million (ppm).

Samples for absorption measurements were deposited on NEGS1 fused-silica substrates (25.4 mm diameter, 6 mm thickness).

\subsection{Ion-beam analysis (IBA)}
The chemical composition of our samples was determined via IBA. We used Rutherford back-scattering spectrometry to detect nitrogen and heavier elements, and elastic recoil detection analysis to detect any possible hydrogen contamination.

\subsubsection{Rutherford back-scattering spectrometry (RBS)}
RBS measurements were performed with two different Van de Graaff particle accelerators, according to their availability in the time frame of our study: the AN2000 accelerator at Laboratori Nazionali di Legnaro (LNL), and the SAFIR facility based on an AN2500 accelerator at the Institut de NanoScience de Paris. In both cases we used a 2.0 MeV 4He+ beam with normal incidence on the samples, and a silicon solid-state detector at a scattering angle of 160 deg (LNL) or 165 deg (SAFIR).

At LNL, we used samples grown on (100) silicon wafers (76.2 mm diameter, 0.5 mm thickness). Relative atomic concentrations and atom areal density of the samples were deduced by matching simulations to the experimental spectra, using a custom-developed code based on the SRIM database \cite{ziegler04}.

At SAFIR, we used samples grown on glassy carbon substrates ($25 \times 25$ mm, 0.5 or 1 mm thickness). Relative atomic concentrations of the samples were deduced by integrating the peaks in the acquired spectra with the SIMNRA software \cite{mayer99,SIMNRA}. We also used SIMNRA to simulate a small background signal observed in the low-energy region of all spectra, and subtracted the results of those simulations from the oxygen and nitrogen contributions. We modeled that background signal as arising from 0.1 to 0.3\% oxygen ambient contamination of the surface of the glassy carbon substrates.

\subsubsection{Elastic recoil detection analysis (ERDA)}
ERDA measurements were performed at the SAFIR facility, on the same samples used for RBS measurements. 
A 2 MeV 4He+ beam was incident on the samples at an angle of 80 deg. The recoiled protons were detected at a forward scattering angle of 30 deg. A 9.5-$\mu$m thick mylar foil placed in front of the detector allowed the elastically recoiled protons into the detector but stopped the high flux of elastically scattered primary beam particles from reaching the detector.

The hydrogen relative atomic concentration of the samples was deduced by matching SIMNRA simulations to the acquired spectra.

\subsection{Mechanical properties}
We estimated the loss angle $\varphi$ of the samples by measuring the frequency and ring-down time of several mechanical resonances \cite{nowick72} of Corning 7980 fused-silica disks (50.0 mm diameter, 0.5 or 1.0 mm thickness), before and after coating deposition, in the 2.5-30 kHz range. In order to avoid systematic damping from suspension and ambient atmospheric pressure, we used a clamp-free, in-vacuum Gentle Nodal Suspension (GeNS) system \cite{cesarini09}. More details about our experimental setup and data analysis can be found in our previous works \cite{granata20b,granata22}.

We derived the volumetric mass density $\rho$ from mass, area and thickness measurements. We measured the mass with an analytical balance, the area was estimated by measuring the diameter of our coated disks with a caliper, and the thickness was inferred from photometric and ellipsometric measurements.

The Young modulus $Y$ and Poisson ratio $\nu$ were estimated by fitting finite-element simulations to data series derived from resonant frequency measurements, according to a method that we developed in earlier works \cite{granata20b}. Further details about our finite-element simulations are available elsewhere \cite{granata16}.

\subsection{Grazing-incidence X-ray diffraction (GI-XRD) and X-ray reflectivity (XRR)}
In order to demonstrate the amorphous phase of the samples, we performed a series of GI-XRD measurements. We used a Philips MRD diffractometer equipped with a Cu tube, operated at 40 kV and 40 mA. The probe beam was collimated and partially mono-chromatized to the Cu K-$\alpha$ line by a parabolic multilayered mirror. The detector was equipped with a parallel-plate collimator as well as with a Soller slit to define the angular acceptance.

With the same diffractometer, by inserting an additional slit in front of the parallel-plate collimator, we performed XRR measurements to estimate the thickness $t$ and density $\rho$ of the samples. Simulations based on the REFLEX software \cite{vignaud19} were fit to the experimental XRR spectra.

For both GI-XRD and XRR measurements, samples were grown on single-side polished (100) silicon substrates (76.2 mm diameter, 0.5 mm thickness).

\subsection{Ar desorption}
As a consequence of the deposition process, the presence of argon in the samples was highly probable. In order to assess whether such contaminant was desorbed upon annealing, X-ray photoelectron spectroscopy and thermal-desorption spectroscopy measurements were performed on samples deposited on single-side polished (100) silicon wafers (76.2 mm diameter, 0.5 mm thickness).

\subsubsection{X-ray photoelectron spectroscopy (XPS)}
XPS experiments were carried out in a ultra-high vacuum setup whose base pressure is usually in the low $10^{-10}$ mbar range. Measurements were performed with an Al K$\alpha$ monochromatic X-ray source (XR50 MF, 1486.6 eV) and with a SPECS PHOIBOS 150 analyzer. The binding energy scale was calibrated on a freshly-sputtered gold foil in electrical contact with the sample, by setting the Au 4f7/2 core-level at 84.0 eV.

\subsubsection{Thermal-desorption spectroscopy (TDS)}
TDS measurements were carried out by placing the samples in the quartz tube of an MTI GSL100 vacuum furnace. The tube was connected to a Pfeiffer Vacuum QMS200 mass spectrometer, its base pressure was of the order of 10$^{-7}$ mbar. The furnace heating rate was set at 180 $^{\circ}$C/hour, and surveys were performed from ambient temperature up to the maximum temperature of 850 $^{\circ}$C allowed by the setup.

\subsection{Coating Brownian thermal noise (CTN)}
CTN measurements were performed in a folded Fabry-Perot cavity where the folding mirror was the high-reflection sample under investigation, using co-resonating second-order orthogonal transverse laser modes TEM02 and TEM20 \cite{gras17}. This technique allows for the routine measurement of thermal noise spectra in the frequency band from $\sim$30 Hz to $\sim$3 kHz, as well as for an alternative estimation of samples' optical absorption via the measurement of the cavity response to thermal load \cite{amato21}.

Samples for CTN measurements were deposited on Corning 7980 fused-silica substrates (25.4 mm diameter, 6 mm thickness).

\section{Results}
The measured properties of our thin-film samples are described in this section.

\subsection{Mitigation of water vapor contamination}
In the early phases of this work, ever since the very first characterizations were performed, an evident anomaly was observed in the acquired photometric spectra. We interpreted it as a consequence of a gradient of refractive index $n$ along the direction of the samples' thickness, with $n$ always decreasing from the surface towards the substrate. Under this hypothesis, a gradient as high as 27\% was required to explain the measured data.

This observation was further confirmed by later ellipsometric measurements, when we also realized that this phenomenon could likely be due to a high value of residual water partial pressure in our coating facility. Indeed, this latter was initially not controlled and in the $10^{-6}-10^{-5}$ mbar range. Results found in the literature also supported the hypothesis of water vapor contamination \cite{bouchier86,burdovitsin83}.

Thus we performed a series of measurements of Fourier-transform infrared spectroscopy (FTIR), Raman spectroscopy, XPS (also coupled to sputtering along the direction of sample thickness, achieved via focused ion beam etching), RBS and ERDA. All those measurements, not shown here for conciseness, revealed an abundant presence of oxygen and hydrogen within the samples, whose relative atomic concentrations were as high as 34\% and 5\%, respectively. Incidentally, we were able to verify that the nitrogen-to-silicon ratio N/Si was equal to or less than 1.0 in all those samples, including those deposited using a pure N$_2$ sputtering beam. This clearly indicated the presence of excess silicon, only partly oxidized, which may cause optical absorption.

In order to suppress the water vapor contamination, we defined the following pre-deposition protocol: the deposition chamber was pumped at least for 12 hours, while the substrate holder was heated up to 100 $^{\circ}$C during the first 3 hours. Thus we were able to obtain a base pressure lower than $2 \times 10^{-7}$ mbar before deposition, and we did not observe any anomaly in the samples' photometric spectra anymore.

A second campaign of FTIR, XPS, RBS, ERDA, and additionally secondary-ion mass spectrometry measurements confirmed that samples deposited after implementing that protocol featured substantially lower concentrations of oxygen and hydrogen, that is, less than 3\% and 0.5\%, respectively. However, it was not be possible to further decrease contamination, as that would have required a base pressure lower than $2 \times 10^{-8}$ mbar \cite{bouchier86}. Such a value could not be achieved in our facility, at that time.

With the only exception of sample 20043, all the results presented in this work were obtained from samples where the refractive index gradient was strongly suppressed, that is, $\Delta n/n<1$\%.

\subsection{Optical properties}

\subsubsection{Refractive index and extinction}

\begin{figure}
\includegraphics[width=0.9\textwidth]{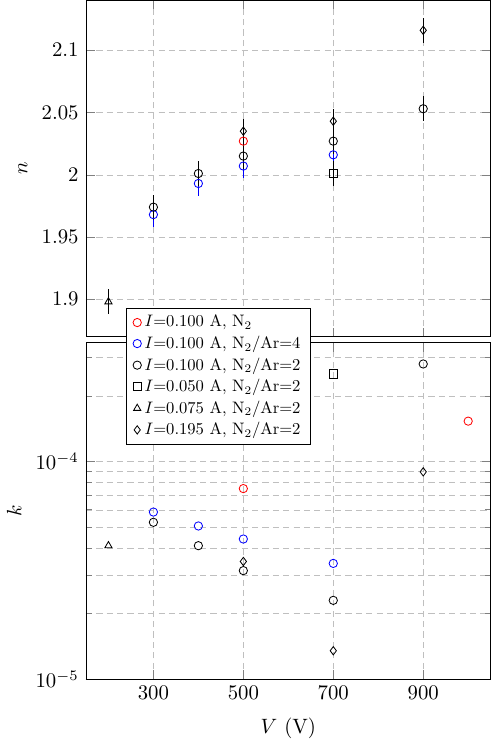}
\caption{\label{FIG_n_k}Measured refractive index $n$ (upper panel) and extinction coefficient $k$ (lower panel) at 1064 nm of IBS SiN$_x$ thin-film samples, as a function of ion beam voltage $V$ and current $I$, and ratio N$_2$/Ar of gas mass flows. Uncertainties on $k$ are $\leq 1$\% (not shown).}
\centering
\end{figure}

\begin{table}
\centering
\begin{threeparttable}
\caption{\label{TAB_n_k} Measured refractive index $n$ and extinction coefficient $k$ at 1064 nm of IBS SiN$_x$ thin-film samples, as a function of ion beam voltage $V$ and current $I$, and Ar and N$_2$ gas mass flows fed to the ion source. Uncertainties on $k$ are $\leq 1$\% (not shown).}
\begin{tabular}{c|cccccc}
\hline \hline
    Sample & $V$ (V) & $I$ (mA) & Ar (sccm) & N$_2$ (sccm) & $n$ & $k \times 10^5$\\ \hline
    20043 & 1000 & 100 & 0 & 10 & \footnotesize{(*)} & 15.3\\
    21012 & 400 & 100 & 5 & 10 & 2.00 $\pm$ 0.01 & 4.1\\
    21013 & 500 & 100 & 5 & 10 & 2.02 $\pm$ 0.01 & 3.2\\
    21014 & 300 & 100 & 5 & 10 & 1.97 $\pm$ 0.01 & 5.3\\
    21016 & 200 & 75 & 5 & 10 & 1.90 $\pm$ 0.01 & 4.1\\
    21017 & 400 & 100 & 3 & 12 & 1.99 $\pm$ 0.01 & 5.1\\
    21018 & 500 & 100 & 3 & 12 & 2.01 $\pm$ 0.01 & 4.4\\
    21019 & 300 & 100 & 3 & 12 & 1.97 $\pm$ 0.01 & 5.9\\
    21020 & 700 & 100 & 3 & 12 & 2.02 $\pm$ 0.01 & 3.4\\
    21021 & 700 & 100 & 5 & 10 & 2.03 $\pm$ 0.01 & 2.3\\
    21022 & 900 & 100 & 5 & 10 & 2.05 $\pm$ 0.01 & 28.0\\
    21023 & 700 & 50 & 5 & 10 & 2.00 $\pm$ 0.01 & 25.2\\
    21025 & 700 & 195 & 5 & 10 & 2.04 $\pm$ 0.01 & 1.4\\
    22013 & 900 & 195 & 5 & 10 & 2.12 $\pm$ 0.01 & 9.0\\
    22014 & 500 & 195 & 5 & 10 & 2.04 $\pm$ 0.01 & 3.5\\
    22015 & 500 & 100 & 0 & 10 & 2.03 $\pm$ 0.01 & 7.5\\
\hline \hline
\end{tabular}
\begin{tablenotes}
    \item[(*)] Not defined.
\end{tablenotes}
\end{threeparttable}
\end{table}
Table \ref{TAB_n_k} lists the refractive index $n$ and extinction coefficient $k$ of the samples at a wavelength $\lambda=1064$ nm, as a function of the deposition parameters. Values of $n$ were obtained from acquired photometric spectra, altogether with values of thickness $t$, whereas the extinction was estimated as follows. Using the known values of $n$ and $t$, a model of each sample was established, based on thin-film theory \cite{furman92}, where $k$ was a free parameter. Every model was thus adjusted to match the values of absorbance $A$ measured at 1064 nm via PTD, expressed in ppm. Figure \ref{FIG_n_k} illustrates the results presented in Table \ref{TAB_n_k}.

We found that $n$ was a monotonic increasing function of the ion beam voltage $V$, in agreement with results found in the literature \cite{lee10,ray94}, whereas we observed that the ion beam elemental composition and current $I$ only had a limited effect on it.

The voltage $V$ also had a noticeable effect on the $k$ coefficient, whose minimum was obtained at 700 V, and the same held for the ion beam composition. The lowest $k$ values were obtained for a mass flow ratio N$_2$/Ar equal to 2. This outcome is in contrast with results reported in the literature, indicating that the optimal ratio would rather be 4 \cite{kitabatake86}. Regardless, the largest effect on $k$ was obtained by varying the ion beam current $I$. At $V = 700$ V, we were able to decrease $k$ by more than an order of magnitude, from $2.1 \times 10^{-4}$ to $1.5 \times 10^{-5}$, by increasing $I$ up to 195 mA. That was the highest current value allowed by our experimental setup in order to have a stable plasma in the ion beam source. 

Thus a minimum of extinction was achieved with sample 21025 for $V = 700$ V, $I = 195$ mA and N$_2$/Ar = 2. Since then, we used those same parameters to deposit multiple nominally identical samples and observed repeatable results. For instance, the average extinction coefficient and refractive index of a subset of fifteen samples were $k=(1.5 \pm 0.2) \times 10^{-5}$ and $n = 2.04 \pm 0.01$, respectively, where the uncertainties are standard deviations. After in-air annealing at 900 $^\circ$C for 10 hours, the extinction decreased to $k = (6.4 \pm 0.2) \times 10^{-6}$. Compared to the tantala-titania layers in current mirror coatings of gravitational-wave interferometers, which show $n = 2.09 \pm 0.01$ and $k \sim 10^{-7}$ \cite{granata20b}, the refractive index of sample 21025 is only 2\% lower. Its extinction, however, while being to the best of our knowledge amongst the lowest ever measured on this class of thin films, is still excessively high. Indeed Wallace et al. measured significantly lower extinction values, although this was very likely due to the fact that their non-stoichiometric SiN$_x$ thin films were highly oxidized (oxynitrides) \cite{wallace24}.

\begin{figure}
\includegraphics[width=\textwidth]{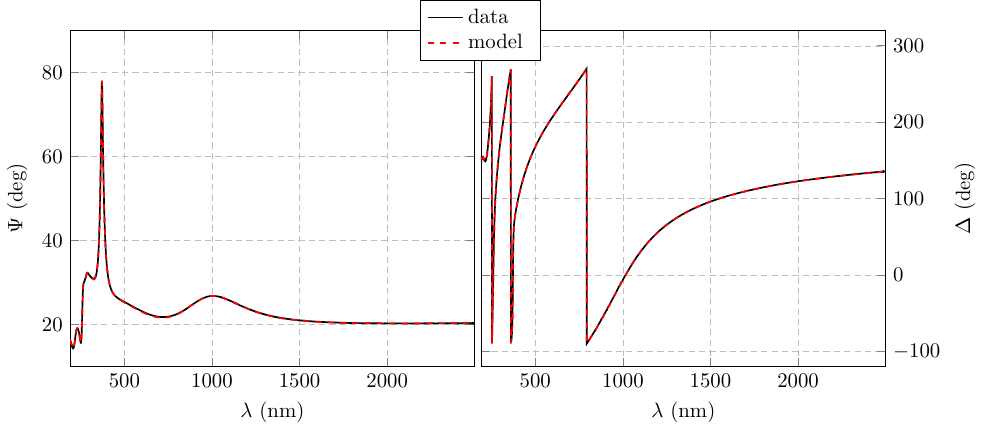}
\caption{\label{FIG_SE_23028}Measured $\Psi$ and $\Delta$ spectra of IBS SiN$_x$ thin-film sample 23028 (acquired at an incidence angle $\theta$ = 60 deg) and their best-fit model, as a function of wavelength $\lambda$.}
\centering
\end{figure}

\begin{figure}
\includegraphics[width=0.9\textwidth]{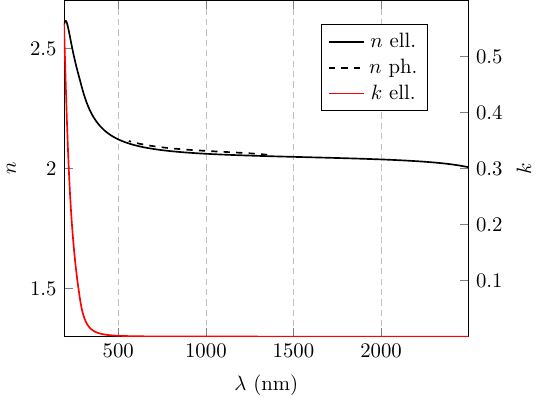}
\caption{\label{FIG_SE_23028_n_k}Refractive index $n$ and extinction coefficient $k$ of IBS SiN$_x$ thin-film sample 23028 as a function of wavelength $\lambda$, as deduced from photometric and ellipsometric spectra. For $\lambda > 560$ nm, $k$ is lower than the sensitivity threshold of the instruments.}
\centering
\end{figure}

Ellipsometric spectra were acquired on samples 21045, 21048, 23017, 23018 and 23028, all deposited using the same set of parameters as sample 21025. By way of example, Fig. \ref{FIG_SE_23028} shows $\Psi$ and $\Delta$ spectra of sample 23028, acquired at an incidence angle $\theta$ = 60 deg, and Fig. \ref{FIG_SE_23028_n_k} shows the dispersion law and the extinction curve we derived from that data. Within their respective measurement uncertainties, refractive index values derived from ellipsometric and photometric measurements and analyses were found to be in agreement. Values of $n$ at 1064, 1550 and 2000 nm are particularly relevant for the operation of current and future gravitational-wave interferometers \cite{aLIGO,AdV,KAGRA,ET} and are listed in Table \ref{TAB_23028_n}, compared against results found in the literature \cite{lee10,wallace24,lambrinos96,lee08,bouchier87}.
\begin{table}
\centering
\caption{\label{TAB_23028_n} Refractive index $n$ of IBS SiN$_x$ thin-film sample 23028 as a function of wavelength $\lambda$, compared to results found in the literature.}
\begin{tabular}{ccccccccc}
\hline\hline
    $\lambda$ (nm) & \multicolumn{2}{c}{$n$}\\
	& This work & Literature\\ \hline
    500 & 2.12 $\pm$ 0.01 & 2.07 \cite{lee10}\\
    546 & 2.11 $\pm$ 0.01 & 2.0 \cite{bouchier87}\\
	633 & 2.09 $\pm$ 0.01 & 2.04 \cite{lambrinos96}\\
	1064 & 2.06 $\pm$ 0.01 & 2.05 \cite{lee08}\\
	1550 & 2.05 $\pm$ 0.01 & 2.01 \cite{wallace24}\\
    2000 & 2.04 $\pm$ 0.01 & \\
\hline\hline
\end{tabular}
\end{table}

The band gap of sample 23028 derived from ellipsometric measurement was found to be $3.6 \pm 0.1$ eV, to be compared to the value of 4.2 eV measured on a stoichiometric IBS Si$_3$N$_4$ thin film by Ray et al. \cite{ray94}. According to that study, the lower value of sample 23028 could be ascribed to the presence of excess silicon.

\subsection{Elemental composition}
\begin{figure}
    \centering
    \includegraphics[width=\textwidth]{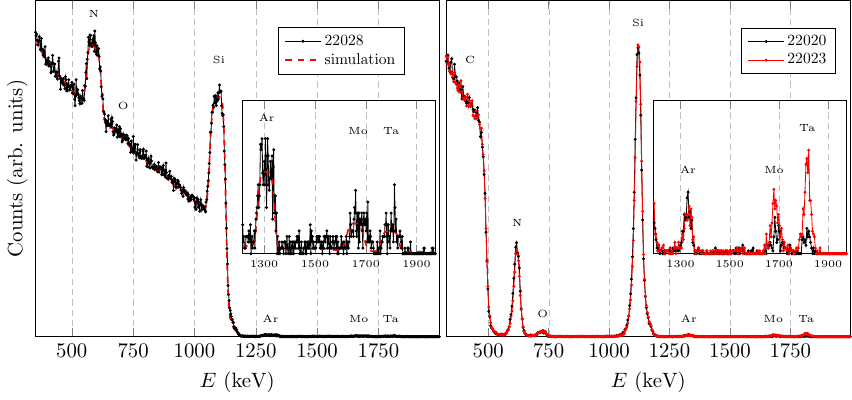}
    \caption{\label{FIG_RBS} Comparison of measured and simulated RBS spectra of IBS SiN$_x$ thin-film sample 22028 deposited on silicon (left panel), and of measured RBS spectra of samples 22020 and 22023 deposited on glassy carbon (right panel). Insets: zoom on the higher-energy spectral region.}
\end{figure}

In order to study more thoroughly the conditions that resulted in minimal optical absorption, we focused our attention on samples deposited on glassy carbon substrates, which allowed to clearly identify and discard the substrate background signal, and using $I = 195$ mA, N$_2$/Ar = 2 and varying $V$. Table \ref{TAB_RBS} presents the relative atomic concentrations of samples 22020, 22022 and 22023, as deduced from acquired RBS spectra. Sample 22020 was deposited using $V = 700$ V as in sample 21025, which featured the lowest optical absorption. Samples 22022 and 22023 were deposited using $V = 500$ V and $V = 900$ V, respectively. By way of example, Fig. \ref{FIG_RBS} shows the measured and simulated RBS spectra of samples 22020 and 22023.
\begin{table}
\centering
\begin{threeparttable}
    \caption{\label{TAB_RBS}Measured relative atomic concentrations of elements (\%) of IBS SiN$_x$ thin films, as a function of ion beam voltage $V$.}
\begin{tabular}{l|cccccccccccc}
    \hline\hline
    Sample & $V$ (V) & H & N & Si & O & Ar & Mo & Ta & N/Si\\ \hline
    22022 & 500 &  & 51.6 & 45.7 & 2.5 & 0.18 & 0.013 & 0.001 & 1.13 &\\
    22020 & 700 & 0.4$^{(*)}$ & 50.4 & 46.4 & 3.0 & 0.17 & 0.014 & 0.004 & 1.08 &\\
    22023 & 900 &  & 48.4 & 48.5 & 2.9 & 0.16 & 0.036 & 0.017 & 1.00 &\\
    \hline\hline
\end{tabular}
\begin{tablenotes}
    \item[(*)] As measured on sample 23017, deposited under identical conditions.
\end{tablenotes}
\end{threeparttable}
\end{table}

Comparative analysis of RBS results of Table \ref{TAB_RBS} first shows that using $I = 195$ mA and N$_2$/Ar = 2 settings slightly increased the N/Si ratio, which reached a value of about 1.1. The beam voltage $V$, on the other hand, only appreciably changed {\color{red} it} for $V = 900$ V, when it equalled 1.0. This seems to contradict the results found in the literature, which indicate that it would be possible to determine a range of $V$ values yielding stoichiometric samples \cite{lee10,huang97,bouchier86,bosseboeuf85,bouchier87}. It should be noted however that these results appear to be in sharp disagreement as to the values to be used. For instance, Bouchier et al. found that stoichiometric samples were obtained using $V \leq 500$ V \cite{bouchier86,bosseboeuf85,bouchier87}, while Huang et al. and Lee et al. found that stoichiometric samples were instead obtained for 700 V $< V <$ 900 V \cite{huang97} or even $V = 1100$ V \cite{lee10}, respectively. Anyway, it remains that, in our specific case, the best deposition conditions identified so far produced sub-stoichiometric, nitrogen-poor samples.

\begin{figure}
    \centering
    \includegraphics[width=0.90\textwidth]{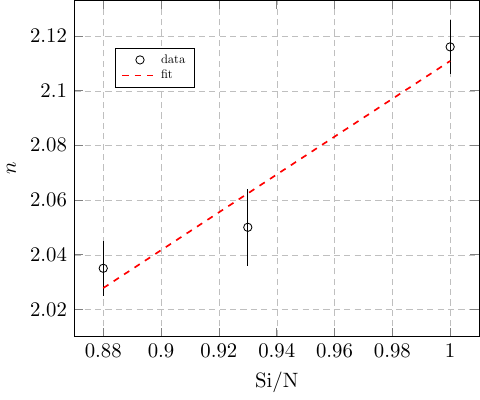}
    \caption{\label{FIG_n_Si-N} Refractive index $n$ of IBS SiN$_x$ thin films at 1064 nm, as a function of their Si/N ratio as measured by RBS. The best-fit slope is shown.}
\end{figure}

By combining data of Tables \ref{TAB_n_k} and \ref{TAB_RBS} altogether, we observed that the refractive index $n$ at 1064 nm varies linearly with the Si/N ratios as measured by RBS. This is shown on Fig. \ref{FIG_n_Si-N}. For the deposition conditions under discussion, we found that
\begin{equation}
n = (0.7 \pm 0.1) \times \text{Si/N} + (1.4 \pm 0.1) ,
\end{equation}
which is the same empirical linear dependence previously observed with different kinds of SiN$_x$ thin films at 633 nm \cite{ray94,jones85,claassen83}.

\subsubsection{Contaminants}
Data of Table \ref{TAB_RBS} also highlights the presence of a variety of contaminants, such as argon, oxygen and hydrogen, molybdenum and tantalum. The persistent presence of oxygen and hydrogen is indeed compatible with the current base pressure inside our coating machine \cite{bouchier86}, which is of the order of $10^{-7}$ mbar and hence determined essentially by the residual partial pressures of water vapor and hydrogen. The presence of argon is most likely due to a fraction of the sputtering particles back-scattered into the growing film \cite{cevro95}, and that of molybdenum can be explained by the fact that the biased output grids of the ion beam source are made of that material. Instead, totally unexpected and yet unexplained is the presence of tantalum. Although we performed a variety of tests, it was not possible for us to isolate the source of such contaminant.

It is noteworthy that the relative concentration of argon is globally independent of the beam voltage $V$ and equal to $\sim0.2$\%, i.e., a factor of ten lower than what observed in previous studies \cite{huang97}, whereas those of molybdenum and tantalum seem only weakly correlated to it. Furthermore, by examining Tables \ref{TAB_n_k} and \ref{TAB_RBS} altogether, it can be deduced that the samples'  optical absorption clearly depends on $V$ but is only loosely correlated to their N/Si ratio or to their relative concentration of Mo and Ta.

\begin{figure}
    \centering
    \includegraphics[width=0.90\textwidth]{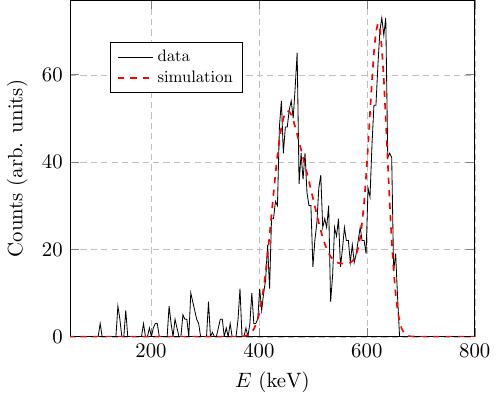}
    \caption{\label{FIG_ERDA} Measured and simulated ERDA spectra of IBS SiN$_x$ thin-film sample 23017, deposited on glassy carbon.}
\end{figure}

Figure \ref{FIG_ERDA} shows the measured and simulated ERDA spectra of sample 23017, which are representative of those obtained from our other samples. It can be seen there that, at first glance,  the measured relative concentration of hydrogen appeared to be highly inhomogeneous across the sample thickness. However, the two peaks observed at $\sim 453$ and $\sim 630$ keV were likely due to ambient contamination at the substrate and coating surface, respectively. Under this hypothesis, the hydrogen content of sample 23017 should be evaluated in the region in between the peaks, hence amounting to 0.4\%. If the peaks were to be included in the analysis, the relative concentration would indeed increase, up to 1.2\%. The small signal observed below 400 keV was identified as an artifact of the experimental setup and not considered in our analysis.

\subsubsection{Ar desorption}
\begin{figure}
    \centering
    \includegraphics[width=0.90\textwidth]{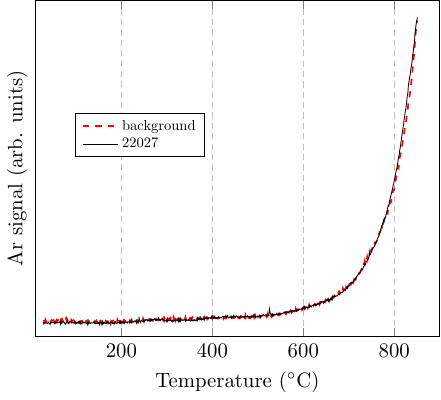}
    \caption{\label{FIG_TDS} Desorption of Ar from IBS SiNx thin-film sample 22027 upon in-vacuum annealing, compared to the background signal of the experimental setup. The heating rate was 180 $^{\circ}$C/h.}
\end{figure}

Figure \ref{FIG_TDS} shows the Ar signal measured by the mass spectrometer during the in-vacuum annealing of sample 22027 ($V=700$ V, $I = 195$ mA, N$_2$/Ar = 2, deposited on a silicon substrate), compared to the background signal of the furnace quartz tube used for the experiment. Both curves overlap, suggesting that the release of Ar from the sample is negligible. The signal increase at higher temperatures is due to unavoidable increase of background pressure inside the furnace. This outcome is different from what was previously observed with IBS metal oxide coatings, which displayed clear desorption peaks \cite{paolone24,lalande24,paolone22,paolone21}.

\begin{figure}
    \centering
    \includegraphics[width=\textwidth]{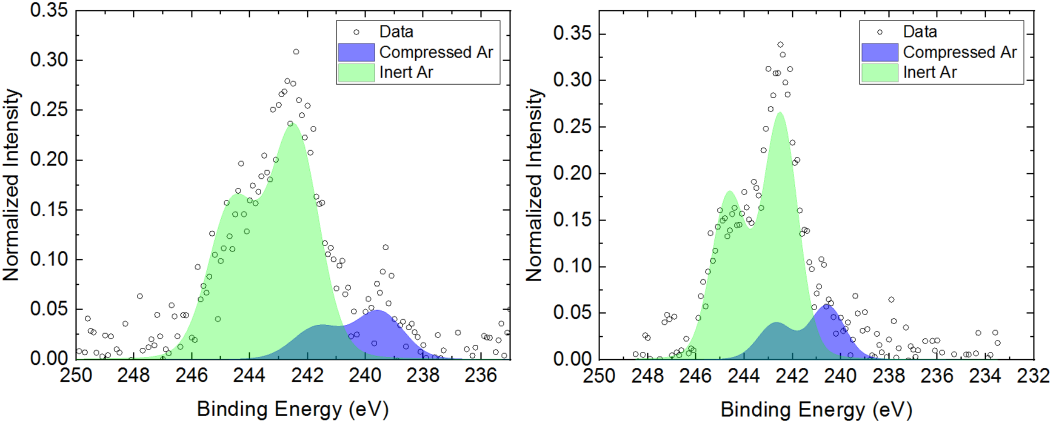}
    \caption{\label{FIG_XPS} Deconvolution of measured XPS Ar 2p spectra of IBS SiNx thin-film sample 22027, before (left) and after (right) in-vacuum annealing at 850 $^{\circ}$C.}
\end{figure}

Figure \ref{FIG_XPS} shows the XPS spectra of the Ar 2p signal normalized with respect to the background noise of sample 22027, acquired before and after the in-vacuum annealing at 850 $^{\circ}$C carried out on the occasion of the TDS experiment. The XPS measurements detected the presence of a relative atomic concentration of Ar of $\sim$0.2\% in the pristine sample, in agreement with the results of RBS measurements. The normalized signal of the sample after annealing turned out to be identical to that acquired before annealing, indicating that Ar was not significantly desorbed and hence confirming the results of TDS measurements. The presence of a component with lower binding energy shifts is typically related to Ar atoms trapped in a more compressed matrix \cite{prenzel12,citrin74}.

\subsection{Mechanical properties}

\subsubsection{Density and phase}
\begin{figure}
    \centering
    \includegraphics[width=0.9\textwidth]{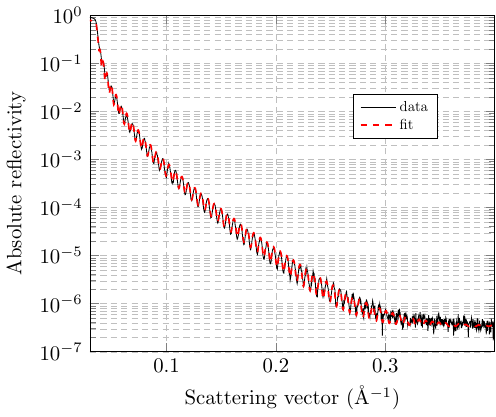}
    \caption{\label{FIG_XRR} Measured XRR spectrum of IBS SiNx thin-film sample 22028. A best-fit REFLEX \cite{vignaud19} simulation is also shown.}
\end{figure}

In order to estimate its density $\rho$, sample 22028 ($V=700$ V, $I = 195$ mA, N$_2$/Ar = 2) was  measured using XRR and RBS. Figure \ref{FIG_XRR} shows the acquired XRR spectrum, together with its best-fit simulation where the sample was modeled as a thin SiN$_x$ layer on top of a Si substrate of semi-infinite thickness. The density value deduced from such simulation was 3.04 $\pm$ 0.01 g/cm$^3$. 
Figure \ref{FIG_RBS} shows the experimental and simulated RBS spectra. In this case, $\rho = 3.038 \pm 0.006$ g/cm$^3$ was calculated using values of atom areal density and relative atomic concentrations, as deduced from stopping power simulations, and of thickness, as deduced from ellipsometric measurements.

A value of density was also derived from mass, area and thickness measurements of sample d2008, deposited under identical conditions. Table \ref{TAB_rho_compar} lists the three density values obtained from independent measurements, showing their agreement. For comparison, reported values for similar IBS SiN$_x$ thin films, crystalline Si$_3$N$_4$ and thermally-grown amorphous Si$_3$N$_4$ are 2.8 $\pm$ 0.28 g/cm$^3$, 3.44 and 3.12 g/cm$^3$, respectively \cite{lambrinos96}.
\begin{table}
\caption{\label{TAB_rho_compar} Density $\rho$ of nominally-identical IBS SiN$_x$ thin films, obtained from three independent measurement methods.}
\begin{tabular}{ccc|cccc}
    \hline\hline
    Sample & \multicolumn{2}{c}{Annealing} & \multicolumn{3}{c}{$\rho$ (g/cm$^3$)}\\
    &  &  & XRR & RBS & Mass/Volume\\ \hline
    22028 &  &  & 3.04 $\pm$ 0.01 & 3.038 $\pm$ 0.006 \\
    d2008 &  &  &  &  & 3.0 $\pm$ 0.5 \\
    22028 & 1000 $^\circ$C & 10 h & $3.06 \pm 0.01$\\
    \hline\hline
\end{tabular}
\end{table}

\begin{figure}
    \centering
    \includegraphics[width=0.9\textwidth]{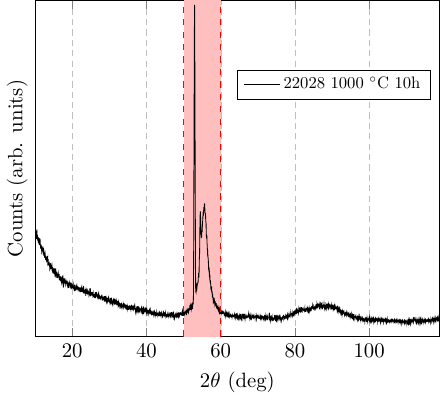}
    \caption{\label{FIG_GIXRD} Measured GI-XRD diffractogram of IBS SiN$_x$ thin-film sample 22028 deposited on silicon, as measured after in-air annealing at 1000 $^\circ$C for 10 hours. Diffraction peaks between 50 and 60 deg (red shaded area) are due to the substrate background signal.}
\end{figure}

Sample 22028 was then treated in-air at 1000 $^\circ$C for 10 hours and used again in RBS and XRR measurements, to observe the effect of annealing. We found that annealing induced a slight increase of the sample density to $3.06 \pm 0.01$ g/cm$^3$, and at the same time an appreciable increase of the oxygen relative concentration, up to a value compatible with the formation of a 7-nm thick layer of stoichiometric SiO$_2$ at the sample surface. The full oxidation of the topmost layer at the sample surface was found to be in agreement with results found in the literature \cite{fourrier91}. As shown by the diffractogram presented on Fig. \ref{FIG_GIXRD}, the phase of the sample remained amorphous after annealing.

Finally, Table \ref{TAB_rho} presents the density $\rho$ of samples deposited using different source parameters, as estimated from mass and volume measurements.
\begin{table}
\caption{\label{TAB_rho} Density $\rho$ of IBS SiN$_x$ thin films as a function of ion beam voltage $V$ and current $I$, and Ar and N$_2$ gas mass flows, as estimated from mass and volume measurements.}
\begin{tabular}{c|cccccc}
    \hline\hline
    Sample & $V$ (V) & $I$ (mA) & Ar (sccm) & N$_2$ (sccm) & $\rho$ (g/cm$^3$)\\ \hline
    c21013 & 700 & 100 & 5 & 10 & 3.1 $\pm$ 0.2\\
    c21015 & 1000 & 75 & 0 & 10 & 2.9 $\pm$ 0.2\\
    d2008 & 700 & 195 & 5 & 10 & 3.0 $\pm$ 0.5\\
    \hline\hline
\end{tabular}
\end{table}

\subsubsection{Coating loss angle}
\begin{figure}
    \centering
    \includegraphics[width=0.9\textwidth]{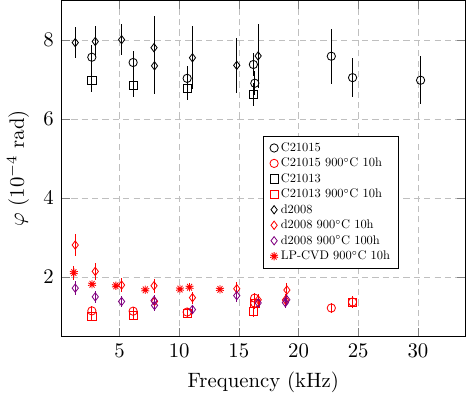}
    \caption{\label{FIG_phi_c} Measured loss angle $\varphi$ of IBS SiN$_x$ thin films deposited using different values of ion beam voltage $V$ and current $I$, and ratio N$_2$/Ar of gas mass flows, as a function of frequency, before (black markers) and after in-air annealing at 900 $^\circ$C (red and purple markers). Also shown is data of a sample deposited by low-pressure chemical-vapor deposition (LP-CVD), after annealing.}
\end{figure}

Figure \ref{FIG_phi_c} and Table \ref{TAB_coatLoss} show the measured loss angle $\varphi$ of samples c21013, c21015 and d2008, produced using different sets of deposition parameters. Measurements were performed before and after in-air annealing at 900 $^\circ$C, and as a function of frequency. First we tested the effect of the ion beam composition, changing it from pure nitrogen to N$_2$/Ar = 2, then we used two different values of ion beam current $I$ which yielded significantly lower optical absorption, $I=100$ and $I=195$ mA (see Table \ref{TAB_n_k}). For comparison, Fig. \ref{FIG_phi_c} also shows a data set of a sample deposited by low-pressure chemical-vapor deposition (LP-CVD)\footnote{For the LP-CVD sample, the following deposition parameters were used: NH$_3$ mass flow of 60 sccm, SiH$_2$Cl$_2$ mass flow of 20 sccm, working pressure of 1.33 mbar, substrate temperature of 800 $^\circ$C.}.
\begin{table}
\caption{\label{TAB_coatLoss} Loss angle $\varphi$ of IBS SiN$_x$ thin films, as measured at $\sim$2.8 kHz, as a function of ion beam voltage $V$ and current $I$, Ar and N$_2$ gas mass flows, and annealing. Values of tantala-titania layers in current mirror coatings of Advanced LIGO and Advanced Virgo are presented, for comparison \cite{granata20b}.}
\begin{tabular}{c|cccccccc}
    \hline\hline
    Sample & $V$ (V) & $I$ (mA) & Ar (sccm) & N$_2$ (sccm) & \multicolumn{2}{c}{Annealing} & $\varphi$ ($10^{-4}$ rad) \\ \hline
    c21013 & 700 & 100 & 5 & 10 &  &  & 7.0 $\pm$ 0.3\\
    c21013 & 700 & 100 & 5 & 10 & 900 $^\circ$C & 10 h & 1.0 $\pm$ 0.1\\
    c21015 & 1000 & 75 & 0 & 10 &  &  & 7.6 $\pm$ 0.3\\
    c21015 & 1000 & 75 & 0 & 10 & 900 $^\circ$C & 10 h & 1.2 $\pm$ 0.1\\
    d2008 & 700 & 195 & 5 & 10 & & & 8.0 $\pm$ 0.4\\
    d2008 & 700 & 195 & 5 & 10 & 900 $^\circ$C & 10 h & 2.2 $\pm$ 0.2\\
    d2008 & 700 & 195 & 5 & 10 & 900 $^\circ$C & 100 h & 1.5 $\pm$ 0.2\\
    LP-CVD &  &  &  &  & 900 $^\circ$C & 10 h & 1.8 $\pm$ 0.1\\
    Ta$_2$O$_5$-TiO$_2$ &  &  &  &  & 500 $^\circ$C & 10 h & 3.3 $\pm$ 0.1\\
    \hline\hline
\end{tabular}
\end{table}

We found that the different deposition conditions had a limited impact on the loss angle of the as-deposited samples, which fell within a narrow range of 6.6 to $8 \times 10^{-4}$ rad for all samples. Minimum coating loss angle was obtained with sample c21013, deposited using N$_2$/Ar = 2, $I=100$ mA and $V=700$ V. The spread increased after annealing, at low frequency, over a range of 1 to $2.8 \times 10^{-4}$ rad. A minimum was obtained for samples c21013 and c21015, whose data sets were nearly superimposed. Sample d2008, which was deposited using the parameters yielding the lowest optical absorption, showed a higher loss angle. However, it was possible to substantially decrease it by increasing its annealing time from 10 to 100 hours.

At approximately 2.8 kHz, which was the lowest resonant frequency accessible with our samples, we found that the loss angle of samples d2008 and c21013 after annealing is 2.2 and 3.3 times lower than that of tantala-titania layers in current mirror coatings of gravitational-wave interferometers \cite{granata20b}, respectively, and appreciably lower than results found in the literature \cite{wallace24,erler80}. To the best of our knowledge, the loss angle value measured on sample c21013 after annealing is the lowest ever measured on IBS SiN$_x$ thin films.

\subsubsection{Young's modulus and Poisson's ratio}
\begin{figure}
    \centering
    \includegraphics[width=0.9\textwidth]{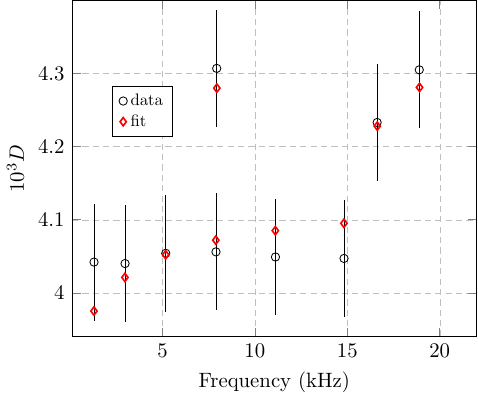}
    \caption{\label{FIG_D} Best-fit finite-element simulation of dilution factor $D$ data derived from resonant frequency measurements of IBS SiN$_x$ thin film sample d2008. This method of analysis is explained in full detail in \cite{granata20b}.}
\end{figure}

Finite-element simulations were fit to data series of dilution factor $D$, derived from resonant frequency measurements \cite{granata20b}. Data series used in this analysis were acquired on samples c21013, c21015 and d2008, produced using different sets of deposition parameters. By way of example, Fig. \ref{FIG_D} illustrates the fit quality for data of sample d2008.
\begin{table}
\caption{\label{TAB_elConst} Young's modulus $Y$ and Poisson's ratio $\nu$ of IBS SiN$_x$ thin films, as a function of ion beam voltage $V$ and current $I$, and Ar and N$_2$ gas mass flows.}
\begin{tabular}{c|cccccc}
    \hline\hline
    Sample & $V$ (V) & $I$ (mA) & Ar (sccm) & N$_2$ (sccm) & $Y$ (GPa) & $\nu$\\ \hline
    c21013 & 700 & 100 & 5 & 10 & 254 $\pm$ 4 & 0.24 $\pm$ 0.04\\
    c21015 & 1000 & 75 & 0 & 10 & 242 $\pm$ 2 & 0.24 $\pm$ 0.02\\
    d2008 & 700 & 195 & 5 & 10 & 256 $\pm$ 2 & 0.25 $\pm$ 0.02\\
    \hline\hline
\end{tabular}
\end{table}

Fit results are presented in Table \ref{TAB_elConst}. The Young's modulus and Poisson's ratio of the samples seem to be rather independent of the deposition parameters, and similar to what can be found in the literature \cite{fletcher18,retajczyk80}.

\subsection{Mirror coatings}
\label{SEC_HR}
The use of multiple materials at different depths in a mirror coating, or {\it multi-material design}, is a design approach which allows to use thin layers of lower Brownian noise despite their higher optical absorption \cite{yam15,steinlechner15}. We thus used our IBS SiN$_x$ thin films with lowest absorption and lower loss angle ($V=700$ V, $I = 195$ mA, N$_2$/Ar = 2) to design and realize a thickness-optimized ternary Bragg reflector, that is, a high-reflection mirror coating composed of stacked thin layers of three different materials \cite{pierro21}. The other materials used were tantala-titania (higher loss angle, lower absorption) and silica (lower loss angle and absorption), as in current mirror coatings of gravitational-wave interferometers \cite{granata20b}. 

\begin{figure}
    \centering
    \includegraphics[width=\textwidth]{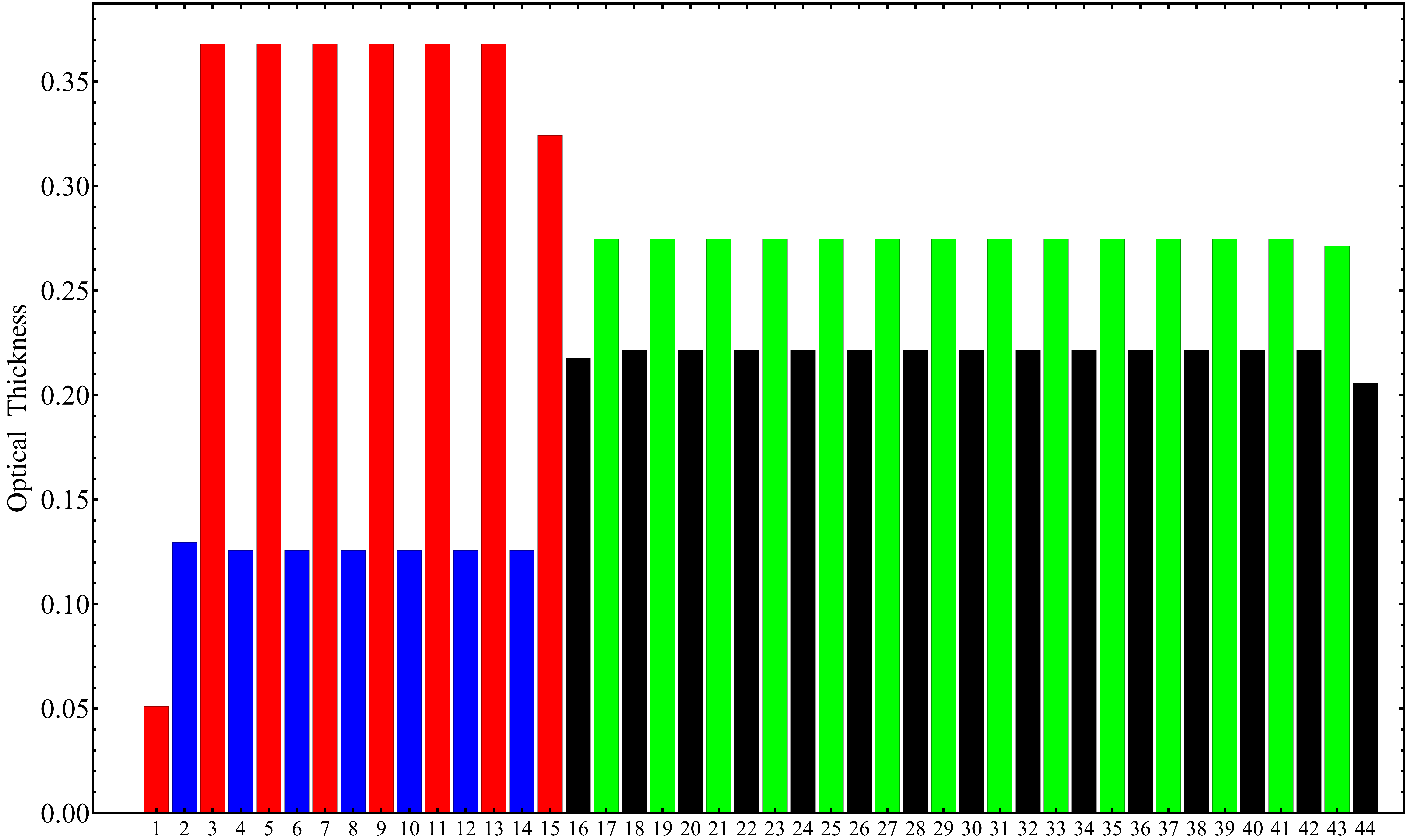}
    \caption{\label{FIG_design} Layout of a thickness-optimized ternary Bragg reflector for operation at $\lambda_0=1064$ nm, composed of IBS SiN$_x$ (black), tantala-titania (blue) and silica (green and red) thin layers. The optical thickness of each layer, normalized to $\lambda_0$, is shown as a function of the layer number (the substrate is set to be on the right, by convention).}
\end{figure}

Our ternary design was found using a custom-developed optimization code to minimize coating transmittance and thermal noise at the same time \cite{pierro21}, based on a multi-objective evolutionary algorithm \cite{hadka13}. For the result to be fairly compatible with the specifications of current mirror coatings in gravitational-wave interferometers \cite{degallaix19,pinard17}, we imposed the constraints of a working wavelength of 1064 nm, a maximum allowed transmittance of 5.6 ppm, and a maximum allowed optical absorption of 1.5 ppm. The resulting design is shown on Fig. \ref{FIG_design}.

Samples of our ternary mirror coating were deposited on Corning 7980 fused silica substrates (25.4 mm diameter, 6 mm thickness), then annealed and measured for transmittance $T$, optical absorption at 1064 nm, and coating thermal noise. All resulted compliant with the target $T$ value of less than 5.6 ppm. By way of example, Table \ref{TAB_CTN} lists the results of coating noise and absorption of sample 10039 as deposited and of sample 10036 after annealing. This latter features a noise amplitude at 100 Hz that is 25\% lower than that of the current coatings of Advanced LIGO and Advanced Virgo. Such measured noise amplitude was found to be in agreement with the prediction of our design optimization code. Concerning the optical absorption, the average value of 1.6 ppm obtained by combining CTN and PTD results was also found to be in agreement with the prediction of our design optimization code.

Work is ongoing to characterize the scattering properties of our ternary mirror coating prototype.
\begin{table}
\begin{tabular}{c|cccc}
\hline\hline
Sample & CTN ($10^{-18}$ m Hz$^{-1/2}$) & a$_{\text{CTN}}$ (ppm) & a$_{\text{PTD}}$ (ppm)\\ \hline
10039 & $(16.2 \pm 0.3) \times \left(\frac{100 \text{Hz}}{f}\right)^{0.48 \pm 0.02}$ & 9.1 $\pm$ 0.3 & 7.6 $\pm$ 0.4\\
10036 & $(10.3 \pm 0.2) \times \left(\frac{100 \text{Hz}}{f}\right)^{0.50 \pm 0.03}$ & 1.3 $\pm$ 0.5 & 1.9 $\pm$ 0.2\\ 
ETM & $(13.7 \pm 0.3) \times \left(\frac{100 \text{Hz}}{f}\right)^{0.45 \pm 0.02}$ & 0.5 $\pm$ 0.2 & 0.27 $\pm$ 0.07\\ \hline\hline
\end{tabular}
\caption{Measured properties of ternary mirror coating samples: coating noise CTN, optical absorption a$_{\text{CTN}}$ as measured in the CTN setup, optical absorption a$_{\text{PTD}}$ as measured via photo-thermal deflection. Sample 10039 was measured as deposited, sample 10036 was measured after annealing. Updated values of current ETM mirror coatings of Advanced LIGO and Advanced Virgo are presented, for comparison.}
\label{TAB_CTN}
\end{table}

\section{Conclusions}
We conducted a broad campaign of production and characterization of IBS silicon nitride thin films, in order to identify the conditions that could lead to minimize optical loss and Brownian noise at the same time.

We found that our films presented systematic deviation from stoichiometry, as well as diverse contaminants. In particular, a very low base pressure was necessary to obtain homogeneous films with oxygen and hydrogen content of less than 3 at.\% and 0.5 at.\%, respectively, likely due to the difficulty in fully removing residual water from the deposition chamber. At such a low base pressure, we tested different deposition parameters (voltage, current intensity and composition of the sputtering ion beam) and annealing conditions, and observed their effects on the elemental composition and optical and mechanical properties of the films.

The ion beam voltage thus appeared to have a measurable effect on the N/Si ratio, which however remained systematically $\leq 1.1$ (that is, the films are nitrogen-deficient) and decreased very moderately with $V$. Voltage was also found to correlate with the refractive index $n$, in a way suggesting that $n$ is primarily controlled by the N/Si ratio. The extinction $k$ due to optical absorption, in contrast, depends quite noticeably on both the beam current $I$ and voltage $V$, and is a non-monotonic function of $V$, which led us to identify the optimal deposition conditions for its minimization. These results should be examined in the light of the observation by Phillip \cite{philipp73} of a systematic, monotonic, dependence of $k$ on the N/Si ratios in thin films deposited by pyrolysis: if both $k$ and $n$ were monotonic functions of N/Si, they should also depend monotonically on each other, which is evidently incompatible with our data. Since we observed that $n$ is a linear function of N/Si, we should conclude that the value of $k$ depends on other factors, as yet unidentified.

In contrast with optical properties, the loss angle $\varphi$ did not appear to significantly depend on the deposition parameters under the explored conditions.

Our study also evidenced a marked effect of high-temperature annealing on both optical absorption and loss angle. Indeed, after annealing at 900 $^\circ$C in air, absorption decreased by a factor $\sim$2, and $\varphi$ by a factor of 4 to 7. We thus managed to obtain IBS SiN$_x$ thin films showing a loss angle and extinction coefficient as low as $(1.0 \pm 0.1) \times 10^{-4}$ rad at $\sim$2.8 kHz and $(6.4 \pm 0.2) \times 10^{-6}$ at 1064 nm, respectively. For this class of thin films, such loss angle is the lowest ever measured, and such optical absorption is amongst the lowest observed \cite{wallace24}. Furthermore, under the hypothesis of a power-law frequency dependence for the loss angle \cite{granata20b}, we might extrapolate to an even lower value of $0.8 \times 10^{-4}$ rad at 100 Hz, in the most sensitive frequency band of gravitational-wave interferometers. However, compared to the extreme requirements of gravitational-wave interferometers, an extinction coefficient of the order of $10^{-6}$ is still excessively high.

Nevertheless, we used our SiN$_x$ thin films to realize a multi-material mirror coating suitable for gravitational-wave interferometers. That coating showed a thermal noise amplitude of $(10.3 \pm 0.2) \times 10^{-18}$ m Hz$^{-1/2}$ at 100 Hz, which is 25\% lower than in current mirror coatings of the Advanced LIGO and Advanced Virgo interferometers, and an optical absorption as low as $(1.9 \pm 0.2)$ parts per million at 1064 nm.

\section*{Acknowledgments}
This work has been promoted by the LMA and partially supported by the Master Projet Couches Minces Optiques of CNRS Nucl\'eaire \& Particules and by the European Gravitational Observatory in the framework of the Advanced Virgo+ project.

The CTN experiment is supported by the LIGO Laboratory. LIGO was constructed by the California Institute of Technology and Massachusetts Institute of Technology with funding from the National Science Foundation and operates under cooperative agreement PHY-1764464. CTN data analysis and figure generation were carried out using MATLAB software provided by The MathWorks, Inc.

M. Magnozzi gratefully acknowledges the RADIATE Transnational Access grant n. 824096 (proposal 21002507-ST) funded by the Horizon 2020 research and innovation programme of the European Union, for access to the SAFIR ion beam facility at the Institut des NanoSciences de Paris.

The integrated XPS measurements were carried out at the SmartLab facility of the Department of Physics at Sapienza Universit\`a di Roma.

We thank I. Roch and G. Cochez of the Institut d'Electronique, de Micro\'electronique et de Nanotechnologie for providing an LP-CVD sample, in the framework of the French RENATECH network. 

In the document repositories of the LIGO and Virgo Collaborations, this work has numbers LIGO-P2400299 and VIR-0630C-24, respectively.

\section*{Data availability}
The data that support the findings of this article are available from the authors upon reasonable request.


\begin{thebibliography}{99}
\bibitem{aLIGO} J. Aasi {\it et al.} (The LIGO Scientific Collaboration), Advanced LIGO, {\it Class. Quantum Grav.} {\bf 32}, 074001 (2015).
\bibitem{AdV} F. Acernese {\it et al.} (The Virgo Collaboration), Advanced Virgo: a second-generation interferometric gravitational wave detector, {\it Class. Quantum Grav.} {\bf 32}, 024001 (2015).
\bibitem{KAGRA} Y. Aso, Y. Michimura, K. Somiya, M. Ando, O. Miyakawa, T. Sekiguchi, D. Tatsumi, and H. Yamamoto (The KAGRA Collaboration), Interferometer design of the KAGRA gravitational wave detector, {\it Phys. Rev. D} {\bf 88}, 043007 (2013).
\bibitem{saulson17} P. R. Saulson, {\it Fundamentals of Interferometric Gravitational Wave Detectors} (World Scientific, Singapore, 2017).
\bibitem{degallaix19} J. Degallaix, C. Michel, B. Sassolas, A. Allocca, G. Cagnoli, L. Balzarini, V. Dolique, R. Flaminio, D. Forest, M. Granata, B. Lagrange, N. Straniero, J. Teillon, and L. Pinard, Large and extremely low loss: the unique challenges of gravitational wave mirrors, J. Opt. Soc. Am. A 36, C85 (2019).
\bibitem{pinard17} L. Pinard, C. Michel, B. Sassolas, L. Balzarini, J. Degallaix, V. Dolique, R. Flaminio, D. Forest, M. Granata, B. Lagrange, N. Straniero, J. Teillon, and G. Cagnoli, Mirrors used in the LIGO interferometers for first detection of gravitational waves, Appl. Opti. 56, C11 (2017).
\bibitem{levin98} Yu. Levin, Internal thermal noise in the LIGO test masses: A direct approach, Phys. Rev. D 57, 659 (1998).
\bibitem{saulson90} P. R. Saulson, Thermal noise in mechanical experiments, Phys. Rev. D 42, 2437 (1990).
\bibitem{villar10} A. E. Villar, E. D. Black, R. DeSalvo, K. G. Libbrecht, C. Michel, N. Morgado, L. Pinard, I. M. Pinto, V. Pierro, V. Galdi, M. Principe, and I. Taurasi, Measurement of thermal noise in multilayer coatings with optimized layer thickness, {\it Phys. Rev. D} {\bf 81}, 122001 (2010).
\bibitem{granata20b} M. Granata, A. Amato, L. Balzarini, M. Canepa, J. Degallaix, D. Forest, V. Dolique, L. Mereni, C. Michel, L. Pinard, B. Sassolas, J. Teillon, and G. Cagnoli, Amorphous optical coatings of present gravitational-wave interferometers, {\it Class. Quantum Grav.} {\bf 37}, 095004 (2020).
\bibitem{mori23} Y. Mori, Y. Nakayama, K. Yamamoto, T. Ushiba, D. Forest, C. Michel, L. Pinard, J. Teillon, and G. Cagnoli, TiO$_2$ doping effect on reflective coating mechanical loss for gravitational wave detection at low temperature, Phys. Rev. D 109, 102008 (2024).
\bibitem{ET} The Einstein Telescope Steering Committee, {\it Einstein Telescope design report update}, technical note \href{https://apps.et-gw.eu/tds/ql/?c=15418}{ET-0007B-20}, (2020).
\bibitem{callen52} H. B. Callen and R. F. Greene, On a Theorem of Irreversible Thermodynamics, Phys. Rev. 86, 702 (1952).
\bibitem{penn03} S. D. Penn, P. H. Sneddon, H. Armandula, J. C. Betzwieser, G. Cagnoli, J. Camp, D. R. M. Crooks, M. M. Fejer, A. M. Gretarsson, G. M. Harry, J. Hough, S. E. Kittelberger, M. J. Mortonson, R. Route, S. Rowan, and C. C. Vassiliou, Mechanical loss in tantala/silica dielectric mirror coatings, Class. Quantum Grav. 20, 2917 (2003).
\bibitem{granata20a} M. Granata, A. Amato, G. Cagnoli, M. Coulon, J. Degallaix, D. Forest, L. Mereni, C. Michel, L. Pinard, B. Sassolas, and J. Teillon, Progress in the measurement and reduction of thermal noise in optical coatings for gravitational-wave detectors, Appl. Opt. 59, A229 (2020).
\bibitem{mcghee23} G. I. McGhee, V. Spagnuolo, N. Demos, S. C. Tait, P. G. Murray, M. Chicoine, P. Dabadie, S. Gras, J. Hough, G. A. Iandolo, R. Johnston, V. Martinez, O. Patane, S. Rowan, F. Schiettekatte, J. R. Smith, L. Terkowski, L. Zhang, M. Evans, I. W. Martin, and J. Steinlechner, Titania Mixed with Silica: A Low Thermal-Noise Coating Material for Gravitational-Wave Detectors, Phys. Rev. Lett. 131, 171401 (2023).
\bibitem{vajente21} G. Vajente, L. Yang, A. Davenport, M. Fazio, A. Ananyeva, L. Zhang, G. Billingsley, K. Prasai, A. Markosyan, R. Bassiri, M. M. Fejer, M. Chicoine, F. Schiettekatte, and C. S. Menoni, Low Mechanical Loss TiO$_2$:GeO$_2$ Coatings for Reduced Thermal Noise in Gravitational Wave Interferometers, Phys. Rev. Lett. 127, 071101 (2021).
\bibitem{lee10} C.-C. Lee et al., Reduction of residual stress in optical silicon nitride thin films prepared by radio-frequency ion beam sputtering deposition, Opt. Eng. 49, 063802 (2010).
\bibitem{liu07} X. Liu, T. H. Metcalf, Q. Wang, and D. M. Photiadis, Elastic Properties of Several Silicon Nitride Films, MRS Online Proceedings Library 989, 2201 (2006).
\bibitem{pan18} H.-W. Pan, L.-C. Kuo, S.-Y. Huang, M.-Y. Wu, Y.-H. Juang, C.-W. Lee, H.-C. Chen, T. T. Wen, and S. Chao, Silicon nitride films fabricated by a plasma-enhanced chemical vapor deposition method for coatings of the laser interferometer gravitational wave detector, Phys. Rev. D 97, 022004 (2018).
\bibitem{steinlechner17} J. Steinlechner, C. Kr\"{u}ger, I. W. Martin, A. Bell, J. Hough, H. Kaufer, S. Rowan, R. Schnabel, and S. Steinlechner, Optical absorption of silicon nitride membranes at 1064 nm and at 1550 nm, Phys. Rev. D 96, 022007 (2017).
\bibitem{wallace24} G. S. Wallace, M. Ben Yaala, S. C. Tait, G. Vajente, T. McCanny, C. Clark, D. Gibson, J. Hough, I. W. Martin, S. Rowan, and S. Reid, Non-stoichiometric silicon nitride for future gravitational wave detectors, Class. Quantum Grav. 41, 095005 (2024).
\bibitem{tsai22} D.-S. Tsai, Z.-L. Huang, W.-C. Chang, and S. Chao, Amorphous silicon nitride deposited
by an NH3-free plasma enhanced chemical vapor deposition method for the coatings of the next generation laser interferometer gravitational waves detector, Class. Quantum Grav. 39, 15LT01 (2022). 
\bibitem{lee08} K.-H. Lee et al., SiNx optical thin films prepared by RF ion-beam sputtering and residual stress elimination technique, Proc. SPIE 7067, Advances in Thin-Film Coatings for Optical Applications V, 70670K (2008).
\bibitem{huang97} L. Huang et al., Structure and composition studies for silicon nitride thin films deposited by single ion beam sputter deposition, Thin Solid Films 299, 104 (1997).
\bibitem{lambrinos96} M. F. Lambrinos, R. Valizadeh, J. S. Colligon, Effects of bombardment on optical properties during the deposition of silicon nitride by reactive ion-beam sputtering, Appl. Opt. 35, 3620 (1996).
\bibitem{ray94} S. K. Ray et al., Effect of reactive ion bombardment on the properties of silicon nitride and oxynitride films deposited by ionbeam sputtering, J. Appl. Phys. 75, 8145 (1994).
\bibitem{bouchier86} D. Bouchier, A. Bosseboeuf, In situ Auger electron spectroscopy investigation of the chemical bonding of ion-beam-deposited silicon nitride, Thin Solid Films 139, 95 (1986).
\bibitem{bosseboeuf85} A. Bosseboeuf, D. Bouchier, Semi-quantitative in situ Auger analysis of silicon nitride layers deposited by reactive ion beam sputtering, Surf. Sci. 162, 695 (1985).
\bibitem{grill83} A. Grill, Ion beam reactively sputtered silicon nitride coatings, Vacuum 33, 329 (1983).
\bibitem{bouchier83} D. Bouchier et al., Low temperature deposition of silicon nitride by reactive ion-beam sputtering, J. Electrochem. Soc. 130, 638 (1983).
\bibitem{erler80} H. J. Erler, G. Reisse, C. Weissmantel, Nitride film deposition by reactive ion beam sputtering, Thin Solid Films 65, 233 (1980).
\bibitem{weissmantel76} C. Weissmantel, Reactive film preparation, Thin Solid Films 32, 11 (1976).
\bibitem{bouchier91} D. Bouchier, A. Bosseboeuf, Angular distributions of sputtered atoms for low-energy nitrogen irradiation of silicon, Appl. Phys. A 53, 179 (1991).
\bibitem{burdovitsin83} V. A. Burdovitsin, Silicon nitride and oxynitride films prepared by ion beam reactive sputtering, Thin Solid Films 105, 197 (1983).
\bibitem{fazio20} M. A. Fazio, G. Vajente, A. Ananyeva, A. Markosyan, R. Bassiri, M. M. Fejer, C. S. Menoni, Structure and morphology of low mechanical loss TiO2-doped Ta2O5, Opt. Mater. Express 10, 1687 (2020).
\bibitem{cisneros98} J. I. Cisneros, Optical characterization of dielectric and semiconductor thin films by use of transmission data, Appl. Opt. 37, 5262 (1998).
\bibitem{fujiwara07} H. Fujiwara, Spectroscopic Ellipsometry: Principles and Applications (Wiley, Hoboken, 2007).
\bibitem{ferlauto02} A. Ferlauto et al., Analytical model for the optical functions of amorphous semiconductors from the near-infrared to ultraviolet: applications in thin film photovoltaics, J. Appl. Phys. 92, 2424 (2002).
\bibitem{amato19} A. Amato et al., Optical properties of high-quality oxide coating materials used in gravitational-wave advanced detectors, J. Phys. Mater. 2, 035004 (2019).
\bibitem{amato20} A. Amato et al., Observation of a correlation between internal friction and Urbach energy in amorphous oxides thin films,  Sci. Rep. 10, 1670 (2020).
\bibitem{boccara80} A. C. Boccara, D. Fournier, W. Jackson, and N. M. Amer, Sensitive photothermal deflection technique for measuring absorption in optically thin media, {\it Opt. Lett.} {\bf 5}, 377 (1980).
\bibitem{ziegler04} J. F. Ziegler, M. D. Ziegler, J. P. Biersack, Nuclear instruments and methods in physics research section B: Beam interactions with materials and atoms, SRIM-2003 219, 1027 (2004).
\bibitem{mayer99} M. Mayer, {\it SIMNRA, a simulation program for the analysis of NRA, RBS and ERDA} in {\it AIP Conference Proceedings} {\bf 475} (AIP, College Park, 1999), p. 541.
\bibitem{SIMNRA} SIMNRA website: https://mam.home.ipp.mpg.de
\bibitem{nowick72} A. Nowick and B. Berry, {\it Anelastic Relaxation in Crystalline Solids} (Academic Press, New York, 1972), p. 582-602.
\bibitem{cesarini09} E. Cesarini, M. Lorenzini, E. Campagna, F. Martelli, F. Piergiovanni, F. Vetrano, G. Losurdo, and G. Cagnoli, A ``gentle'' nodal suspension for measurements of the acoustic attenuation in materials, {\it Rev. Sci. Instrum.} {\bf 80}, 053904 (2009).
\bibitem{granata22} M. Granata, A. Amato, M. Bischi, M. Bazzan, G. Cagnoli, M. Canepa, M. Chicoine, A. Di Michele, G. Favaro, D. Forest, G. M. Guidi, G. Maggioni, F. Martelli, M. Menotta, M. Montani, F. Piergiovanni, and F. Schiettekatte, Optical and mechanical properties of ion-beam-sputtered MgF$_2$ thin films for gravitational-wave interferometers, Phys. Rev. Applied 17, 034058 (2022).
\bibitem{granata16} M. Granata, E. Saracco, N. Morgado, A. Cajgfinger, G. Cagnoli, J. Degallaix, V. Dolique, D. Forest, J. Franc, C. Michel, L. Pinard, and R. Flaminio, Mechanical loss in state-of-the-art amorphous optical coatings, {\it Phys. Rev. D} {\bf 93}, 012007 (2016).
\bibitem{vignaud19} G. Vignaud, A. Gibaud, Reflex: A program for the analysis of specular x-ray and neutron reflectivity data, J. Appl. Crystallogr. 52, 201 (2019).
\bibitem{gras17} S. Gras, H. Yu, W. Yam, D. Martynov, and M. Evans, Audio-band coating thermal noise measurement for Advanced LIGO with a multimode optical resonator, Phys. Rev. D 95, 022001 (2017).
\bibitem{amato21} A. Amato, G. Cagnoli, M. Granata, B. Sassolas, J. Degallaix, D. Forest, C. Michel, L. Pinard, N. Demos, S. Gras, M. Evans, A. Di Michele, and M. Canepa, Optical and mechanical properties of ion-beam-sputtered Nb$_2$O$_5$ and TiO$_2-$Nb$_2$O$_5$ thin films for gravitational-wave interferometers and an improved measurement of coating thermal noise in Advanced LIGO, Phys. Rev. D 103, 072001 (2021).
\bibitem{furman92} S. A. Furman, and A. V. Tikhonravov, {\it Basics of Optics of Multilayer Systems} (Editions Fronti{\`e}res, Gif-sur-Yvette, 1992), p. 22-46.
\bibitem{kitabatake86} M. Kitabatake, K. Wasa, Hydrogen‐free SiN films deposited by ion beam sputtering, Applied Physics Letters 49, 927 (1986).
\bibitem{bouchier87} D. Bouchier, A. Bosseboeuf, G. Gautherin, Interface properties of silicon nitride deposited by ion beam sputtering on silicon, Proceedings of the Symposium on silicon nitride and silicon dioxide thin insulating films, edited by V. J. Kapoor and K. T. Hankins (The Electrochemical Society Inc., Pennington, New Jersey), 527 (1987).
\bibitem{jones85} B. L. Jones, High stability, plasma deposited, amorphous silicon nitride for thin film transistors, J. Non-Cryst. Solids 77-78, 957 (1985).
\bibitem{claassen83} W. A. P. Claassen, W. G. J. N. Valkenburg, F. H. P. M. Habraken, and Y. Tamminga, Characterization of Plasma Silicon Nitride Layers, J. Electrochem. Soc. 130, 2419 (1983).
\bibitem{cevro95} M. Cevro and G. Carter, Ion-beam and dual-ion-beam sputter deposition of tantalum oxide films, Opt. Eng. 34, 596 (1995).
\bibitem{paolone24} A. Paolone, M. Bazzan, G. Favaro, F. Borondics, G. Nemeth, F. Capitani, S. Swaraj, R. Belkhou, R. Flammini, J. Teillon, C. Michel, D. Hofman, M. Granata, Argon release, crystallization, morphological and optical changes of ion-beam sputtered Ta2O5 thin films during thermal treatments, Heliyon 11, e42009 (2025).
\bibitem{lalande24} E. Lalande, A. Davenport, L. Marchand, A. Markosyan, D. Martinez, A. Paolone, M. Rezac, M. Bazzan, M. Chicoine, J.L. Colaux, M. Coulon, M.M. Fejer, A.W. Lussier, E. Majorana, L. Martinu, C. Menoni, C. Michel, F. Ricci, F. Schiettekatte, N. Shcheblanov, J.R. Smith, J. Teillon, G. Terwagne, G. Vajente, Ar transport and blister growth kinetics in titania-doped germania-based optical coatings. Class. Quantum Grav. in press https://doi.org/10.1088/1361-6382/ad3ffb.
\bibitem{paolone22} A. Paolone, E. Placidi, E. Stellino, M. G. Betti, E. Majorana, C. Mariani, A. Nucara, O. Palumbo, P. Postorino, M. Sbroscia, F. Trequattrini, M. Granata, D. Hofman, C. Michel, L. Pinard, A. Lemaitre, N. Shcheblanov, G. Cagnoli and F. Ricci, Argon and other defects in amorphous SiO2 coatings for gravitational-wave detectors, Coatings 12, 1001 (2022).
\bibitem{paolone21} A. Paolone, E. Placidi, E. Stellino, M.G. Betti, E. Majorana, C. Mariani, A. Nucara, O. Palumbo, P. Postorino, I. Rago, F. Trequattrini, M. Granata, J. Teillon, D. Hofman, C. Michel, A. Lemaitre, N. Shcheblanov, G. Cagnoli, F. Ricci, Effects of the annealing of amorphous Ta2O5 coatings produced by ion beam sputtering concerning the effusion of argon and the chemical composition, J. Non-Cryst. Solids 557, 120651 (2021).
\bibitem{prenzel12} M. Prenzel, T. de los Arcos, A. Kortmann, J. Winter and A. von Keudell, Embedded argon as a tool for sampling local structure in thin plasma deposited aluminum oxide films, J. Appl. Phys. 112, 103306 (2012).
\bibitem{citrin74} P. H. Citrin and D. R. Hamann, Measurement and calculation of polarization and potential-energy effects on core-electron binding energies in solids: X-ray photoemission of rare gases implanted in noble metals, Phys. Rev. B 10, 4948 (1974).
\bibitem{fourrier91} A. Fourrier, A. Bosseboeuf, D. Bouchier, and G. Gautherin, Thermal Oxidation in Wet Oxygen of Reactive Ion‐Beam Sputter‐Deposited Silicon Nitride Films, J. Electrochem. Soc. 138, 1084 (1991).
\bibitem{fletcher18} M. Fletcher, S. Tait, J. Steinlechner, I. W. Martin, A. S. Bell, J. Hough, S. Rowan, R. Schnabel, Effect of Stress and Temperature on the Optical Properties of Silicon Nitride Membranes at 1550 nm, Front. Mater. 5 (2018).
\bibitem{retajczyk80} T. F. Retajczyk, A.K. Sinha, Elastic stiffness and thermal expansion coefficients of various refractory silicides and silicon nitride films, Thin Solid Films 70, 241 (1980).
\bibitem{yam15} W. Yam, S. Gras, and M. Evans, Multimaterial coatings with reduced thermal noise, Phys. Rev. D 91, 042002 (2015).
\bibitem{steinlechner15} J. Steinlechner, I. W. Martin, J. Hough, C. Kr\"uger, S. Rowan, and R. Schnabel, Thermal noise reduction and absorption optimization via multimaterial coatings, Phys. Rev. D 91, 042001 (2015).
\bibitem{pierro21} V. Pierro, V. Fiumara, F. Chiadini, V. Granata, O. Durante, J. Neilson,  C. Di Giorgio, R. Fittipaldi, G. Carapella, F. Bobba, M. Principe, I. M. Pinto, Ternary quarter wavelength coatings for gravitational wave detector mirrors: Design optimization via exhaustive search, 
Phys. Rev. Res. 3, 023172 (2021)
\bibitem{hadka13} D. Hadka and P. Reed, Borg: An Auto-Adaptive Many-Objective Evolutionary Computing Framework, Evol. Comput. 21, 231 (2013).
\bibitem{philipp73} H. R. Philipp, Optical Properties of Silicon Nitride, J. Electrochem. Soc. 120, 295 (1973).
\end{thebibliography}
\end{document}